\documentstyle[12pt]{article}
\newcommand{\spur}[1]{\not\! #1 \,}
\newcommand{\be}{\begin{equation}}
\newcommand{\dd}{\displaystyle}
\newcommand{\ee}{\end{equation}}
\newcommand{\bea}{\begin{eqnarray}}
\newcommand{\eea}{\end{eqnarray}}
\newcommand{\abs}[1]{\left\vert#1\right\vert}
\begin{document}
\hfill$\vcenter{
 \hbox{\bf BARI-TH 410/01}
\hbox{\bf UGVA-DPT-2001-01-1093} }$
\begin{center}
{\Large\bf\boldmath {Effective Field Theory for the }}\vskip0.3cm
{\Large\bf\boldmath{Crystalline Colour Superconductive Phase of}}
\vskip0.3cm {\Large\bf\boldmath{QCD}}
\\ \rm \vskip1pc {\large
R. Casalbuoni$^{a,b}$,  R. Gatto$^c$, M. Mannarelli$^{c,d,e}$
and\\ G. Nardulli$^{d,e}$}\\ \vspace{5mm} {\it{$^a$Dipartimento di
Fisica, Universit\`a di Firenze, I-50125 Firenze, Italia
\\
$^b$I.N.F.N., Sezione di Firenze, I-50125 Firenze, Italia\\
$^c$D\'epart. de Physique Th\'eorique, Universit\'e de Gen\`eve,
CH-1211 Gen\`eve 4, Suisse\\ $^d$Dipartimento di Fisica,
Universit\`a di Bari, I-70124 Bari, Italia  \\$^e$I.N.F.N.,
Sezione di Bari, I-70124 Bari, Italia }}
\end{center}

\begin{abstract}
We present an effective field theory for high density, low
temperature QCD in the crystalline colour superconductive phase
({\it LOFF} phase). This interesting phase of QCD is characterized
by a gap parameter with  a crystalline pattern, breaking
traslational and rotational invariance, and could have
astrophysical applications. In the effective theory the fermions
have a Majorana mass, which, besides colour, breaks translation
and rotation symmetries. Fermions couple to the  phonon arising
from the breaking of rotation and translation invariance.
Integrating out the fermions leads eventually to an effective
lagrangian in terms of the phonon field only, which satisfies an
anisotropic dispersion relation.
\end{abstract}

\vskip1.cm
\section{Introduction}
We shall consider massless quarks of  three colours and two
different flavours. At zero density the theory is invariant under
the group $SU(3)_c\times  SU(2)_L\times SU(2)_R$. We shall
disregard in the sequel further invariances under $U(1)$, not
essential for the following derivations. The model we shall
consider is based on the approximation: $m_1=m_u\approx
0,~m_2=m_d\approx 0$ and large $m_s$, of the full QCD theory with
three flavours.

At higher densities we introduce two chemical potentials, $\mu_1$
and $\mu_2$ for the two species, with (assuming
$\mu_1\geq\mu_2$)\be \mu_1-\mu_2=\delta
\mu\ll\mu=\frac{\mu_1+\mu_2}2 \; .\ee
  Physical situations with $\delta\mu\neq
0$ are the most likely to occur in nature. For example, the
so-called glitches of neutron stars \cite{LOFF} (see also
\cite{LOFF3}) should be studied  in this hypothesis. More
generally, it is  theoretically important, independently of
applications, to study in detail each single phase of the QCD
phase-diagram, which implies the analysis for generic values of
$\delta\mu$.  This note is devoted to the construction of an
effective theory for the physically interesting case
$\delta\mu\neq 0$ . Before describing our formalism, let us start
however with a brief  review of the  $\delta \mu = 0$ case.

QCD in the limit $\delta \mu=0,~\mu\to\infty,~T=0$ has been first
studied in \cite{alford} and subsequently by a number of authors
\cite{others}, who have shown that a phenomenon of colour
superconductivity takes place. Effects of this phase should most
likely be seen in astrophysical contexts, for instance in
connection with magnetic fields in quark matter \cite{alford1},
quark star oscillations \cite{masden}, cooling of neutron stars
\cite{page}, and supernova neutrinos \cite{carter}.

The origin of the superconducting (BCS) phase  can be traced back
to the vacuum expectation value of the operator \be
\psi^T_{i\alpha }C\psi_{j\beta }\; ,\ee ($\alpha,\, \beta=1,2,3$
colour indices; $i,\,j=1,2$ flavour indices) that, for $\mu$
sufficiently large, is  non vanishing.   If $\psi$ is used to
represent a left-handed two-component Weyl spinor and
$C=i\sigma_2$, one gets indeed:
\be\epsilon^{\alpha\beta\gamma}\epsilon_{ij}<\psi^T_{i\alpha
}C\psi_{j\beta}>=\Delta\delta^{\gamma 3}\; .\label{lab2}\ee The
right handed field satisfies the same relation with $\Delta\to -
\Delta$ (we assume $\Delta>0$). The condensate (\ref{lab2}) breaks
the original symmetry group $SU(3)_c\otimes SU(2)_L\otimes
SU(2)_R$ down to \be SU(2)_c\times SU(2)_L\times SU(2)_R\ .\ee The
chiral group remains unbroken, while the original colour symmetry
group is broken to $SU(2)_c$, with generators $T^A$ corresponding
to the generators $T^1,T^2,T^3$ of $SU(3)_c$. Therefore  three
gluons remain massless whereas the remaining five acquire a mass.
There are no Nambu-Goldstone Bosons. The effective theory to
describe this situation  has been discussed in \cite{sannino} (the
three massless flavor case was studied in \cite{cfleffective}).

The microscopic dynamics results from a mechanism analogous to the
formation of an electron Cooper pair in a BCS superconductor. At
$T=0$ the only QCD interactions are those involving fermions near
the Fermi surface. Quarks inside the Fermi sphere cannot interact
because of the Pauli principle, unless the interactions involve
large momentum exchanges. In this way the quarks can escape the
Fermi surface, but these processes are disfavoured, as large
momentum transfers imply small couplings due to the asymptotic
freedom property of QCD. Even though interactions of fermions near
the Fermi surface involve momenta of the order of $ \mu$, their
effects are not necessarily negligible. As a matter of fact, even
a small attractive interaction between fermions near the Fermi
surface and carrying opposite momenta can create an instability
and give rise to coherent effects. This is what really happens
\cite{alford} and the result is the formation of a diquark
condensate, as expressed by (\ref{lab2}). We stress again that the
only relevant fermion degrees of freedom are therefore those near
the Fermi surface.

This picture holds also for $\delta\mu\neq 0$ provided
$\delta\mu\ll \Delta $. On the other hand, for
$\delta\mu\approx\Delta $, the picture changes significantly. The
analysis in \cite{LOFF} shows that there exist two values of
$\delta\mu$, $\delta\mu_1$ and $\delta\mu_2$, such that,
for\be\delta\mu\in(\delta\mu_1,\delta\mu_2)\ ,\label{loff7} \ee
the high density quark-gluon matter is in a phase characterized by
the breaking of translational and rotational invariance, due to
the presence of a scalar and a vector condensate. This phenomenon
is called crystalline colour superconductivity of QCD and the
relative phase is named {\it LOFF} phase, from the initials of the
authors \cite{LOFF2} who have studied a similar phase in quantum
electrodynamics. The {\it LOFF} phase, is energetically favoured
as compared to the BCS phase if (\ref{loff7}) holds. The authors
of ref. \cite{LOFF} find $\delta\mu_1=0.71\Delta$ and
$\delta\mu_2=0.744\Delta$ for $\mu=0.4~GeV$ and $\Delta=40~MeV$.

The aim of the present note is to present an effective lagrangian
approach to this phase. By integrating out the negative energy
fermion fields and the gluons and introducing velocity-dependent
positive energy quark fields we shall first obtain an effective
theory where the effective quark fields are basically free, but
possess a Majorana mass term breaking translational and rotational
invariance. This term arises from the presence of a scalar and a
vector condensate, as discussed in \cite{LOFF}. This lagrangian
represents an extension of the results of \cite{gatto} to the
crystalline colour superconductive phase. We shall then couple the
theory to  the  field which describes the Nambu-Goldstone boson
arising by the breaking of the above mentioned symmetries. The
integration of the fermion degrees of freedom eventually leads to
an effective theory containing as dynamical fields only the NGB
field.

\section{Effective theory near the Fermi surface}

To start with, we extend the approach of Ref. \cite{gatto} to the
two flavour case. We allow $\delta\mu\neq 0$. If $p_j$ is the
momentum of the quark having flavour $j$, we write: \be p_j=\mu_j
v+\ell_j\ .\label{2.1} \ee Here $v^\mu =(0,\vec v)$, where $\vec
v$ is the Fermi velocity. Since we wish to describe excitations
near the Fermi surface, we shall  limit the functional integration
in the generating functional to fermion fields satisfying
\cite{Hong} \be|\ell_j| < \mu_j \ .\ee Therefore, introducing
fields $\psi_\pm$, corresponding to positive and negative energies
for the massless left-handed Weyl fermion of flavour $j$ in the
chemical potential $\mu_j$,  we have \bea\psi(x)& =& \sum_{\vec
v}e^{-i\mu_jv\cdot x}\int_{\abs{\ell_j} < \mu_j
}\frac{d^4\ell_j}{(2\pi)^4}e^{-i\ell_j\cdot x}\,\psi_{\vec v
}(\ell_j)=\cr
 &=&\sum_{\vec v}e^{-i\mu_jv\cdot x}
 \left[\psi_+(x)+\psi_-(x)\right]\label{decomposition}
 ,\eea where
\be\psi_\pm(x)=\frac{1\pm \vec \alpha\cdot \vec v }2\int_{\abs
\ell < \mu_j } \frac{d^4\ell}{(2\pi)^4}e^{-i\ell\cdot x}
\psi_{\vec v}(\ell)\ .\ee Here $\displaystyle\sum_{\vec v} $ means
an average over the Fermi velocities and
\be\psi_\pm(x)\equiv\psi_{\pm,\vec v}(x) \ee are
velocity-dependent fields. The meaning of this decomposition is
that the negative energy fields $\psi_-$ can be expressed in terms
of the positive energy ones through the formula \be
\psi_-=-\frac{1}{2\mu_j}\gamma_0\spur{\partial}_T\psi_+\ ,\ee
 which  results from the equations of motion, and therefore can be
integrated out. Here
 \bea\spur{\partial}_T
&=&\partial_\mu \gamma_T^\mu\\ \gamma_T^\mu
&=&\gamma_\lambda\left(2 g^{\mu\lambda}-V^\mu\tilde
V^\lambda-V^\lambda\tilde V^\mu \right)\eea and \bea
V^\mu&=&(1,\,+\vec v)\cr \tilde V^\mu&=&(1,\,-\vec v)\ .\eea
Eliminating $\psi_-$ results in an effective theory which at the
lowest order is described by the lagrangian \cite{gatto}:
 \be {\mathcal L}_0\ =\ \sum_{\vec v}\left(\psi_+^\dag i
V^\mu\partial_\mu\psi_+ \ -\ \frac
1{2\mu_j}\psi^\dag_+\left(\spur{\partial}_T\right)^2\psi_+\right)
\ +\ (L\to R)\ .\label{lo}\ee

The term $\propto 1/\mu_j$ is in general non leading and can be
neglected, which we will do in the sequel\footnote{In \cite{gatto}
it has been shown that it contributes to the Meissner gluon mass
in the colour flavour locking phase of QCD \cite{cfl}. Since we
are not interested here in the calculation of the gluon mass
arising from the colour breaking of 2-flavour QCD we can neglect
it altogether; for a more complete discussion on this see
\cite{mannarelli}. We have not included in the previous expression
the gluon fields which play no role in the subsequent
discussion.}. We note explicitly that the fields appearing in this
equation depend on the same velocity $\vec v$ because of the
Riemann Lebesgue lemma. Off-diagonal contributions would have the
form: \be \sum_{ \vec v_1\neq\vec v_2} e^{i\mu_j(\vec v_1-\vec
v_2)\cdot \vec x} \psi_{+,\vec v_2}^\dag i
V\cdot\partial\,\psi_{+,\vec v_1}+...\ee that give negligible
contributions in the large chemical potential limit, due to the
rapid oscillations of the exponential function.

 It is useful to use a different basis for the fermion fields. We
introduce: \bea \psi_{+,i\alpha }&=&
\sum_{A=0}^3\frac{(\sigma_A)_{i\alpha}}{\sqrt 2}\varphi_{+}^A
~~~~~~~~(i,\,\alpha=1,\,2)~\cr \psi_{+,13}&=&\varphi_{+}^4\cr
\psi_{+,23}&=&\varphi_{+}^5\ ,\eea where $\sigma_A$ are the Pauli
matrices for $A=1,2,3$ and $\sigma_0=1$. Here clearly
$\varphi_{+}^A$ are positive energy, velocity dependent fields:
\be \varphi_{+}^A\equiv\varphi_{+,\vec v}^A\ . \ee We also
introduce\be \varphi_{-}^A\equiv\varphi_{+,-\vec v}^A\ .\ee
$\varphi_{\pm }^A$ should not be confused with the positive and
negative energy fields; they are both positive energy fields, but
are relative to opposite velocities.

The lagrangian $ {\mathcal L}_0$ can therefore be written as
follows:
 \be {\mathcal L}_0\ =\ \sum_{\vec v}\sum_{A=0}^5
 \varphi_{+}^{A\dag}( i V\cdot\partial)\,\varphi_{+}^A  \ +\ (L\to
R)\ .\label{lobis} \ee Using the fact that the average over
velocities is symmetric we write:
 \be{\mathcal L}_0 =
 \frac 1 2
  \sum_{\vec v}\sum_{A=0}^5\Big
 (\varphi_{+}^{A\,\dag} (i
V\cdot\partial)\, \varphi_{+}^A+ \varphi_{-}^{A\,\dag}( i \tilde
V\cdot\partial)\,\varphi_{-}^A\Big)\ .\ee
 Introducing now
\begin{equation}
\chi^A\ =\left(\begin{array}{c}
  \varphi_{+}^A  \\ {}\\
  C\varphi_{-}^{A\,*}\ ,
\end{array}\right)
\end{equation}
the lagrangian can be written as follows: \be {\mathcal L}_0\ =\
\frac 1 2 \sum_{\vec v}\sum_{A=0}^5 \chi^{A\,\dag}\left(
\begin{array}{cc}
 i V\cdot\partial\ & 0
\\
0 & i \tilde V\cdot\partial
\end{array}
\right)\chi^A\ .\ee
\section{Crystalline colour superconductive phase} As shown in
\cite{LOFF} the vacuum state is characterized by a non vanishing
expectation value of a lagrangian term breaking translational and
rotational invariance. The appearance of this condensate is a
consequence of the fact that for $\mu_1\neq\mu_2$, and in a given
range of $|\mu_1-\mu_2|$ \cite{LOFF}, the formation of a Cooper
pair with a total momentum\be \vec p_1\,+\,\vec p_2\,=\,2\vec q\ee
 is energetically favoured in comparison with the normal
BCS state, if (\ref{loff7}) is satisfied. There are  actually two
condensates, one with the two quarks in  a spin zero state (scalar
condensate) and the other one characterized by total spin 1
(vector condensate). In the BCS state the quarks forming the
Cooper pair have necessarily $S=0$; as a matter of fact, since the
quarks have opposite momenta and equal helicities, they must be in
an antisymmetric spin state. This not true if the total momentum
is not zero and the two quarks can have both $S=0$ and $S=1$.

The possible form of these condensates is discussed in
\cite{LOFF}; in particular the geometric properties of the
crystalline phase have been assumed to arise from a simple plane
wave behaviour of the condensate $\dd\propto e^{2i\vec q\cdot\vec
x}$, though more complicated structures are possible. We shall
limit our analysis to the simplest case; to begin with, we
consider the lagrangian term relative to the scalar condensate. We
shall write it  as follows: \be {\cal L
}^{(s)}_\Delta=-\frac{\Delta^{(s)}}2\, e^{2i\vec q\cdot\vec x}
\epsilon^{\alpha\beta 3}\epsilon_{ij} \psi_{i\alpha}^T(x)C
\psi_{i\beta}(x)\ -(L\to R)+{\rm h.c.} \label{loff5}~.\ee Here
$\psi(x)$ are positive energy left-handed fermion fields.
 We introduce velocity dependent positive energy fields
$\psi_{+,\,\vec v_i;\,i\alpha}$ having flavour $i$; neglecting the
non-leading negative-energy fields we have: \bea {\cal L
}^{(s)}_\Delta&=&-\frac{\Delta^{(s)}} 2 \, e^{2i\vec q\cdot\vec
x}\times\cr&&\sum_{\vec v_1,\vec
v_2}\epsilon_{ij}\epsilon^{\alpha\beta 3} e^{i\mu(\vec v_1+\vec
v_2)\cdot \vec x\,+\,i\frac{\delta\mu}{2}(\vec v_1-\vec v_2)\cdot
\vec x}\psi_{+,\,\vec v_i;\,i\alpha}(x)C \psi_{+,\,\vec
v_j;\,j\beta}(x)\cr && -(L\to R)+{\rm h.c.}\ . \label{loff6}\eea
Because of the Riemann-Lebesgue theorem, in the $\mu\to\infty$
limit, the only non vanishing term in the sum corresponds to $\vec
v_1+\vec v_2=0$. Putting $\vec v=\vec v_1=-\vec v_2$ and \be
\psi_{+,\,\pm\vec v_i;\,i\alpha}(x)~\equiv~\psi_{\pm\vec
v_i;\,i\alpha}(x)~,
 \ee we
have: \be{\cal L }^{(s)}_\Delta=-\frac{\Delta^{(s)}}{2}\,
e^{2i\vec q\cdot\vec x}\sum_{\vec
v}\epsilon_{ij}\epsilon^{\alpha\beta 3}e^{i\delta\mu\vec v\cdot
\vec x} \psi_{+\vec v;\,i\alpha}(x)C \psi_{-\vec
v;\,j\beta}(x)-(L\to R)+{\rm h.c.}\ . \ee This term can be written
as follows: \be{\cal L }^{(s)}_\Delta=\ -\ \sum_{ \vec
v}\frac{\Delta_{\vec q,\,\vec v }^{(s)}(\vec x)}2
\epsilon_{ij}\epsilon^{\alpha\beta 3} \psi_{+\vec v;\,i\alpha}(x)C
\psi_{-\vec v;\,j\beta}(x)\ , \ee with \be \Delta^{(s)}_{\vec
q,\,\vec v }(\vec x)=\Delta^{(s)} e^{i\delta\mu\vec v\cdot \vec
x}e^{2i\vec q\cdot\vec x}. \ee It is clear that this term violates
translation and rotation invariance, since $\vec q$ is fixed. It
also breaks colour symmetry from $SU(3)_c$ down to $SU(2)_c$.

 The term corresponding to the vector condensate in the lagrangian
  can be written as follows: \be{\cal L
}^{(v)}_\Delta=\ -
 \sum_{\vec v}\frac{\Delta^{(v)}_{\vec
q,\,\vec v }(\vec x)}2\sigma^1_{ij}\epsilon^{\alpha\beta 3}
\psi_{+\vec v;\,i\alpha}(x)C(\vec\alpha\cdot\vec n ) \psi_{-\vec
v;\,j\beta}(x)\ , \ee where \be \Delta^{(v)}_{\vec q,\,\vec v
}(\vec x)=\Delta^{(v)} e^{i\delta\mu\vec v\cdot \vec x}e^{2i\vec
q\cdot\vec x}\, .\ee Here $\vec n=\vec q/|\vec q|$  is the
direction corresponding to the total momentum carried by the
Cooper pair. Also this term, besides colour symmetry, breaks
translation and rotation invariance.

In the basis introduced in the previous section, the effective
lagrangian is
 \bea
  {\mathcal L}&=& {\mathcal L}_0 \ +\ {\mathcal L}^{(s)}_\Delta \
  +\ {\mathcal L}^{(v)}_\Delta\ =\cr&&\cr
&=&\frac 1 2 \sum_{\vec v}\sum_{A=0}^5 \chi^{A\,\dag}\left(
\begin{array}{cc}
 i \delta_{AB}V\cdot\partial\ & \,\Delta_{AB}
\\
\,  \Delta_{AB} & i\delta_{AB} \tilde V\cdot\partial
\end{array}\right)\chi^B\ ,\eea
with \be \Delta_{AB}\ =\ \Delta_{AB}^{(s)}\ +\ (\vec v\cdot\vec
n)\ \Delta_{AB}^{(v)}\ .\ee Correspondingly,  the effective action
in momentum space reads: \be S=\sum_{\vec v}\sum_{A,B=0}^5\frac 1
2 \int \frac{d^4\ell}{(2\pi)^4} \frac{d^4\ell^\prime}{(2\pi)^4}
\chi^{A\dag}(\ell^\prime)D^{-1}_{AB}(\ell^\prime,\ell)\chi_B(\ell)~,
 \ee
 where $D^{-1}_{AB}(\ell^\prime,\ell)$ is the inverse propagator
 given by:
\bea &&D^{-1}_{AB}(\ell^\prime,\ell)=\cr &&\cr&&=\left(
\begin{array}{cc}
V\cdot\ell  \delta_{AB}\, \delta^4(\ell^\prime-\ell)&
\Delta_{AB}\, \delta^4(\ell^\prime-\ell+\delta\mu \, v+2 q)
\\
\Delta_{AB}\,\delta^4(\ell^\prime-\ell-\delta\mu\, v-2q)
 &  \tilde V\cdot\ell
 \delta_{AB}\,
\delta^4(\ell^\prime-\ell)
\end{array}\nonumber
\right)\ . \eea Here we have defined $q^\mu=(0,\vec q)$ and we
have used $\psi_-^TC\vec\alpha\cdot \vec n\psi_+=\vec v\cdot\vec
n\psi_-^TC\psi_+$~.  The matrix $\Delta^{(s)}_{AB}$ is diagonal:
\be \Delta_{AB}^{(s)}= \delta_{AB}\Delta_{A}^{(s)} \ee with \be
\Delta^{(s)}_0=\Delta^{(s)},~~~~~~\Delta^{(s)}_1=~\Delta^{(s)}_2=
~\Delta^{(s)}_3~=-\
\Delta^{(s)},~~~~~~\Delta^{(s)}_4=~\Delta^{(s)}_5=0\ .\ee On the
other hand one has \be \Delta_{AB}^{(v)}\ =\  \Delta^{(v)}
\left(\begin{array}{cccccc}
  0 & 0 & 0 & 1 & 0 & 0 \\
  0 & 0 & i & 0 & 0 & 0 \\
  0 & -i & 0 & 0 & 0 & 0 \\
  1 & 0 & 0 & 0 & 0 & 0 \\
  0 & 0 & 0 & 0 & 0 & 0 \\
  0 & 0 & 0 & 0 & 0 & 0
\end{array}\right)\ .
\ee From these equations one can derive the quark propagator,
given by \bea &&D_{AB}(\ell,\ell^{\prime\prime})=\cr
&&\cr&&=\sum_C\left(
\begin{array}{cc}
\tilde V\cdot\ell\,  \delta_{AC}\,
\delta^4(\ell-\ell^{\prime\prime})& -\Delta_{AC}\,
\delta^4(\ell-\ell^{\prime\prime}+\delta\mu \, v+2 q)
\\
-\Delta_{AC}\,\delta^4(\ell-\ell^{\prime\prime}-\delta\mu\, v-2q)
 &   V\cdot\ell\,
 \delta_{AC}\,
\delta^4(\ell-\ell^{\prime\prime})
\end{array}
\right)\times\cr &&\left((V\cdot \ell)(\tilde V\cdot
\ell)-{\bf\Delta}^2\right)_{CB}^{-1} \eea with $\dd
{\bf\Delta}_{AB}^2=\sum_C\Delta_{AC}\Delta_{CB}$. One can observe
that $\Delta_{4B}=\Delta_{5B}=\Delta_{A4}=\Delta_{A5}=0$.
Therefore the propagator for both  the fields $\chi^{4,5}$ is
simply given by \be D(\ell,\ell^{\prime})=\left(
\begin{array}{cc}
 (V\cdot\ell)^{-1}\delta^4(\ell-\ell^{\prime})&0\\
 0&  (\tilde V\cdot\ell)^{-1}\,
\delta^4(\ell-\ell^{\prime})
\end{array}
\right)~.\ee As for the other four fields $\chi^A$,
$A=0,\cdots,3$, we proceed as follows. Let us  put \be
\Delta_0=\Delta^{(s)},~~\Delta_1=\vec v\cdot\vec
n\,\Delta^{(v)}~;\ee we get \be \Delta_{AB}=
\left(\begin{array}{cccc}
   \Delta_0 & 0 & 0 & \Delta_1\\
   0 & -\Delta_0 & i\Delta_1 & 0\\
   0 & -i\Delta_1 & -\Delta_0 & 0\\
   \Delta_1 & 0 & 0 & -\Delta_0
\end{array}\right)~.\ee We note that the explicit appearance of the factor
 $i$ in the
Majorana mass matrix implies the breaking of charge conjugation
according to its standard definition. However  there is invariance
under the following combined operations:  take the complex
conjugation, and send $\vec v\to -\vec v$, as it follows noticing
that, correspondingly, $\Delta_1\to-\Delta_1$.

The mass matrix can be made diagonal by going from the basis
$\chi^A$ to the new basis $\tilde \chi^A$ defined as follows \bea
\tilde\chi^{0}&=&\cos\theta\,\chi^0+\sin\theta\,\chi^3\cr
\tilde\chi^1&=&\frac 1{\sqrt{2}}(\chi^1-i\chi^2)\cr
\tilde\chi^2&=&\frac 1{\sqrt{2}}(\chi^1+i\chi^2)\cr
\tilde\chi^{3}&=&-\sin\theta\,\chi^0+\cos\theta\,\chi^3 \ ,
\label{basis}\eea with $\dd
\tan\theta\,=\,\frac{\Delta_1}{\Delta_0+\sqrt{\Delta_0^2+\Delta_1^2}}
$. In the new basis, the propagator  for the fields
$\tilde\chi^A$, $A=0,\cdots,3$, is given by \bea
&&D_{AB}(\ell,\ell^\prime)= \left(\begin{array}{cc}
  \tilde V\cdot\ell\, \delta^4(
  \ell-\ell^\prime) & -M_A\delta^4(
  \ell-\ell^\prime+\delta\mu v+2q)\\
  -M_A\delta^4(
  \ell-\ell^\prime-\delta\mu v-2q) & V\cdot\ell \,\delta^4(
  \ell-\ell^\prime)\end{array}\right)\times\cr&&\frac
{\delta_{AB}}{(V\cdot\ell)(\tilde
  V\cdot\ell)-M_A^2}\eea
where\bea M^A&=&(M^0,M^1,M^2,M^3)=\cr&&\cr
&=&\left(+\sqrt{\Delta_0^2+\Delta_1^2},
-(\Delta_0-\Delta_1),-(\Delta_0+\Delta_1),-\sqrt{\Delta_0^2+\Delta_1^2}\right)\
.
 \eea
One can note the mass differences between  these states arising
from the  combined action of $\Delta_0$ and  $\Delta_1$.
\smallskip
\section{Effective lagrangian for the phonon}

Let us consider again the breaking terms given by ${\cal
L}_\Delta^{(s)}$ and ${\cal L}_\Delta^{(v)}$, that is \bea {\cal
L}_\Delta&=&{\cal L}_\Delta^{(s)}+{\cal L}_\Delta^{(v)}=\cr&&\cr
&=&-\frac 1 2\, e^{2i\vec q\cdot\vec x}\sum_{\vec v}e^{i\delta\mu
\vec v\cdot\vec x}\left[\Delta^{(s)}\epsilon_{ij}+\vec v\cdot\vec
n\Delta^{(v)}\sigma^1_{ij}\right]\epsilon^{\alpha\beta
3}\psi_{+\vec v;\,i\alpha}\,C\,\psi_{-\vec
v;\,j\beta}\cr&&\cr&&-(L\to R)\ . \label{ldeltaeff}\eea This
condensate breaks explicitly rotations and translations, however
the discussion of the number of NGB's in the case of space
symmetries is a subtle one due to the particular group structure.
In fact rotations and translations cannot be considered
transformations breaking the symmetries of the theory in an
independent way. This is because a translation plus a rotation is
physically equivalent to a translation. We have two sources of
space-time symmetry  breaking in (\ref{ldeltaeff}), one arising
from the exponential term, which breaks both translation and
rotation invariance, and another one in the vector condensate
(only rotation invariance is broken in this term). These two
breakings are locked as the direction of $\vec q$ in the
exponential term coincides with the direction $\vec n$ of the
vector condensate. Let us discuss the consequences of this
situation more closely.

 Let us first consider spatial rotations and the vector condensate.
 We take the
$z$-axis pointing along the direction of $\vec q$. We define the
three-vector \be (\vec R)_i=R_{i}(\xi_1,\xi_2)=(e^{i(\xi_1
L_1+\xi_2 L_2)/f_R})_{i3}\ee with $\vec L$ the generators of the
rotation group in the spin 1 representation, i.e. \be
(L_i)_{jk}=-i\epsilon_{ijk}\ee and  $\xi_1$, $\xi_2$ are fields.
It follows from the orthogonality of the rotation matrices that
\be |\vec R|^2=1~.\label{r1}\ee Moreover we assume\be \langle \vec
R\rangle_0=\vec n\label{r2}\,.\ee Since the fields $R_{i}$
transforms as a vector under rotations, it follows that the
rotational symmetry is restored by the substitution: \be \vec
v\cdot\vec n\to \vec v\cdot\vec R\ .\label{51}\ee

 Let us now consider the exponential factor
 $\exp(2 i\vec q \cdot\vec x)=\exp(2 iq\vec n \cdot\vec x)$
 in (\ref{ldeltaeff}), which
  breaks both rotational and translational invariance. Notice that
  the factor $\exp(i\delta\mu v\cdot x)$ does not break
  translation and rotation symmetries, since it comes from a field
  redefinition in a lagrangian which was originally invariant.
   As we have discussed, we can deal with rotations by the
substitution $\vec n\to\vec R$, i.e. by the replacement:\be
e^{2iq\vec n\cdot\vec x}\to e^{2i q\vec R\cdot\vec x}\ . \ee As it
stands, however, this factor still breaks the translational
invariance, since, under a translation $\vec x\to \vec
x^{\,\prime}=\vec x+\vec a$ it transforms as \be e^{2iq\vec
R\cdot\vec x}\to e^{2i q\vec R\cdot\vec x+2i q\vec R\cdot\vec a}
\ee In order to make the exponential term invariant  we introduce
therefore a new function $T\equiv T(x)$ such that under
translations \be T\to T-2q\vec a\cdot\vec R\,
;\label{translations}\ee if we write the exponential term as
follows \be e^{i\Phi}=e^{2iq\vec R\cdot\vec x+iT}\, , \ee then it
becomes invariant under both translation and rotations. Let us
consider now the combination \be\Phi=2q\vec R\cdot\vec x+T=2\vec
q\cdot\vec x+\phi/f\, ,\label{56}\ee with \be\phi/f=2q(\vec R-\vec
n)\cdot\vec x +T\, .\label{phi1}\ee The function $\phi$ acts as
the phonon (Nambu-Goldstone boson) field associated to the
breaking of the space symmetry. Under a translation $\vec a$ \be
\phi/f\to\phi/f-2\vec q\cdot\vec a\, ,\label{phi}\ee and under an
infinitesimal rotation $\vec x\to \vec x+\delta\vec x$, \be
\phi/f\to\phi/f-2\vec q\cdot \delta \vec x\, .\label{rotation}\ee
moreover the classical phonon field has to satisfy the condition
of being a small fluctuation, i.e. \be |\phi|\le f\,
.\label{52}\ee

Had we considered only the scalar condensate there would be no
need to discuss the field $\vec R$ any longer, as only the
combination $\Phi$ would be relevant. The presence of the vector
condensate forces to introduce the field $\vec R$, as we have seen
in eq. (\ref{51}); however, the condition (\ref{52}) tells us that
the fields $\vec R$ and $T$ are strictly related. In fact, in
order to maintain $\phi$ small, one has to compensate the
fluctuations of $\vec R$ (that are amplified at large distances by
the factor $\vec x$) with large fluctuations of $T$. This means
that the degrees of freedom described by $\vec R$ are not
independent of $T$. We can describe this dependence by writing \be
\vec R =\frac{\vec \nabla\Phi}{|\vec \nabla\Phi|}\, ,\label{r3}\ee
which satisfies (\ref{r1},\ref{r2}). We note that $\Phi$ is an
auxiliary field  while the dynamical field is $\phi$; as a matter
of fact, $\Phi$ and $\phi$ differ by the vacuum expectation value
of $\Phi$ \be \langle\Phi\rangle_0=2\vec q\cdot\vec
x,~~~\langle\phi\rangle_0=0\, .\ee

 In conclusion, the interaction term with the
NGB field   is contained in \bea {\cal L}_{int}&=&-\frac 1 2\,
e^{i\Phi} \sum_{\vec v}e^{i\delta\mu \vec v\cdot\vec
x}\times\cr&&\cr&\times&\left[\Delta^{(s)}\epsilon_{ij}+\Delta^{(v)}(\vec
v\cdot\vec R )\sigma^1_{ij} \right]\epsilon^{\alpha\beta
3}\psi_{i,\alpha,\vec v}\,C\,\psi_{j,\beta,-\vec
v}\cr&&\cr&-&(L\to R)\ . \label{lextrnal}\eea Notice that we have
neglected the breaking of the colour symmetry, which has been
considered elsewhere \cite{sannino}. At the first order in the
fields one gets the following   three-linear coupling:
 \bea
 {\cal L}_{\phi\psi\psi}&=&-\frac{i\phi}{2f} e^{2i\vec q\cdot\vec x}\sum_{\vec
v}e^{i\delta\mu \vec v\cdot\vec x}\left[\Delta^{(s)}\epsilon_{ij}+
\vec v\cdot\vec n
\Delta^{(v)}\sigma^1_{ij}\right]\epsilon^{\alpha\beta
3}\psi_{i,\alpha,\vec v}\,C\,\psi_{j,\beta,-\vec
v}\cr&&\cr&&-\frac{1}{4fq} e^{2i\vec q\cdot\vec x}\sum_{\vec
v}e^{i\delta\mu \vec v\cdot\vec x} \vec v\cdot \left[ \vec
\nabla\phi-\vec n(\vec n\cdot\vec\nabla\phi)\right]
\Delta^{(v)}\sigma^1_{ij}\epsilon^{\alpha\beta
3}\psi_{i,\alpha,\vec v}\,C\,\psi_{j,\beta,-\vec
v}\cr&&\cr&&-(L\to R)\ .
 \label{Trilineare}\eea

Through a bosonization procedure similar to the one employed in
\cite{gatto}, one can derive an effective lagrangian for  the NGB
field. The effective lagrangian must contain only derivative
terms. Polynomial terms are indeed forbidden by translation
invariance, since $\phi$ is not an invariant field (see eq.
(\ref{phi1})). In order to write the kinetic terms is better to
use the auxiliary field $\Phi$ which behaves as a scalar under
both rotations and translation. To avoid the presence of
polynomial terms  in the phonon lagrangian one has to exclude
polynomial terms in the auxiliary field $\Phi$ as well; therefore
the lagrangian should be constructed only by derivative terms.
However since the expectation value of the gradient of $\Phi$ is
given by $\langle\Phi\rangle_0=2\vec q$, we cannot limit the
expansion in the spatial detivatives of $\Phi$ to any finite
order. A real spatial derivative expansion can be made only for
the phonon field $\phi$. With this in mind the most general
invariant lagrangian will contain a tower of space-derivative
terms:
 \be{\cal
L}(\phi,\partial_\mu\phi)=\frac{f^2}
2\left[\dot\Phi^2-\sum_{n=1}^\infty c_n
(|\vec\nabla\Phi|^{2})^n\right]\ .\label{Phi}\ee
 It can be observed that, because of the definition
of $\vec R$, no other vector is available besides $
\vec\nabla\Phi$ (for instance $\vec
R\cdot\vec\nabla\Phi=|\vec\nabla\Phi|$). In this lagrangian $\Phi$
must be thought as a function of the phonon field $\phi$.

Since \be |\vec\nabla\Phi|^2=4q^2+\frac{4q}{f}\vec
n\cdot\vec\nabla\phi+\frac{1}{f^2}|\vec\nabla \phi|^2\ ,\ee and
similar for higher powers,  at the lowest order in the derivatives
of the phonon field $\phi$ we get, neglecting a constant term:
\be{\cal L}(\phi,\partial_\mu\phi)=\frac{1}{2}\left[\dot\phi^2-
v_\parallel^2|\vec\nabla_\parallel\phi|^2-
v^2\left(4qf\vec\nabla_\parallel\phi+|\vec\nabla\phi|^2\right)
\right]\ , \label{phonon}\ee where $\vec\nabla_\parallel\phi=\vec
n\cdot\vec\nabla\phi$ and $v_\parallel^2$,  $v^2$ are constants.
Notice that the linear term gives rise to a surface contribution.
Also, the lagrangian (\ref{Phi}), expressed in terms of the true
scalar $\Phi$ is rotational and translational invariant. On the
other hand, in order to define a {\it bona fide} field we have to
extract from $\Phi$ its space-dependent vacuum expectation value.
This implies that the phonon field $\phi$ does not behave as a
scalar field under translations and rotations (see eqs.
(\ref{phi}) and (\ref{rotation})). As a consequence, the gradient
expansion made in eq.  (\ref{phonon}) seems to lack rotational
invariance. However this is not really true, since  it is not
difficult to check that in eq. (\ref{phonon}) the sum of the two
terms proportional to $v^2$ is rotational invariant, whereas the
variation of the term proportional to $v_\parallel^2$ is
compensated by the variation of a third order term in the spatial
derivatives. The lack of rotational invariance appears only order
by order in the gradient expansion. This should not be too
surprising. In fact we can draw an analogy with the chiral case.
The analogue of the chiral field $\Sigma$ is here the field
$\Phi$ whereas the pion corresponds to the phonon. Expanding the
chiral lagrangian in terms of the pion field we lack, order by
order, chiral invariance, just because the pion transforms non
linearly under chiral rotations. The same happens in the present
case. Therefore, the physical consequence of the extraction of the
expectation value of $\Phi$ is  an anisotropy in the dispersion
relation for the phonon field.
\section{Conclusions}

We have constructed the effective lagrangian decribing the
expected crystalline colour superconductive phase {\it (LOFF}
phase) in high density and low temperature QCD with two massless
flavors. The {\it LOFF } phase has been considered of possible
astrophysical interest. It certainly deserves theoretical study as
one of the components of the QCD phase diagram.
 The effective lagrangian we have constructed here
 extends previous results \cite{gatto} to the
crystalline  phase. The construction is done by first
 integrating out the gluons and the
negative energy (with respect to the Fermi sphere) fermions and
describing positive energy fermions by velocity-dependent fields.
The effective quarks behave as free fields with a Majorana mass
term, breaking (besides colour, to be dealt with as in
\cite{sannino}) translational and rotational invariance. The
theory is then coupled to the phonon arising from the spontaneous
breaking of  space symmetries. Finally one can integrate over the
fermions leaving as dynamical fields only the phonon field.
\vskip.5cm\noindent {\bf Acknowledgement} We would like to thank
Krishna Rajagopal for enlightening advices which have helped us
very much in giving the final form to this paper.


\begin{thebibliography}{1-99}

\bibitem{LOFF}M. Alford, J. A. Bowers and K. Rajagopal,
hep-ph/0008208.

\bibitem{LOFF3} K. Rajagopal, Acta Phys. Polon. {\bf B31} (2000) 3021,
hep-ph/0009058, M. Alford, J. A. Bowers and K. Rajagopal, J.
Phys. {\bf G27} (2001) 541, hep-ph/0009357, T. Sch\"afer and E.
Shuryak, nucl-th/0010049; J. A. Bowers, J. Kundu, K. Rajagopal
and E. Shuster, hep-ph/0101067.

\bibitem{alford}M. Alford, K. Rajagopal and F. Wilczek, Phys.
Lett.B {\bf 422} (1998) 247.
\bibitem{others} For a complete review of the subject and a list
of references see: K. Rajagopal and F. Wilczek, hep-ph/0011333.
For earliest papers on colour superconductivity see B. Barrois,
Nucl. Phys. {\bf B129} (1977) 390, S. Frautschi, Proceedings of
workshop on hadronic matter at extreme density, Erice 1978. See
also: D. Bailin and A. Love, Phys. Rept. {\bf 107} (1984) 325, and
references therein.

\bibitem{alford1} M. Alford, J. Berges and K. Rajagopal,
Nucl.\ Phys.\  {\bf B571}, 269 (2000), hep-ph/9910254.
\bibitem{masden}J. Madsen, Phys. Rev. Lett. {\bf 85}, 10 (2000)
\bibitem{page} D. Page, M. Prakash, J. Lattimer and A. Steiner, Phys. Rev. Lett.
{\bf 85} (2000) 2048, hep-ph/0005094.
\bibitem{carter}G. Carter and S. Reddy, Phys. Rev. {\bf D62} (2000) 103002,
hep-ph/0005228.

\bibitem{sannino}R. Casalbuoni, Z. Duan and F. Sannino,
Phys. Rev. {\bf D62} (2000) 094004, hep-ph/0004207.

\bibitem{cfleffective} R. Casalbuoni and R. Gatto, Phys. Lett. {\bf B464} (1999)
111,  hep-ph/9908227.

\bibitem{LOFF2}A. I. Larkin and Yu. N. Ovchinnikov, Zh. Eksp.
Teor. Fiz. {\bf 47} (1136) 1964 ( Sov. Phys. JETP {\bf 20} (1965)
762); P.Fulde and R. A. Ferrell, Phys. Rev. {\bf 135} (1964) A550.

\bibitem{gatto}R. Casalbuoni, R. Gatto and G. Nardulli,  Phys. Lett.
{\bf B498} (2001) 179, hep-ph/0010321.
\bibitem{Hong} D.~K. ~Hong, Phys. Lett. {\bf B473} (2000) 118,
hep-ph/9812510; D.~K.~ Hong, Nucl. Phys. {\bf B582} (2000) 451,
hep-ph/9905523; S. R. Beane, P. F. Bedaque and M. J. Savage,
Phys.Lett. {\bf B483} (2000) 131, hep-ph/0002209.


\bibitem{cfl}  D.T. Son and M.A.
Stephanov, Phys. Rev. {\bf D61} 074012 (2000), hep-ph/9910491,
erratum: Phys. Rev. {\bf D62} 059902(E)  (2000), hep-ph/0004095.
\bibitem{mannarelli}R. Casalbuoni, R. Gatto, M. Mannarelli
and G. Nardulli, in preparation.
\end{thebibliography}
\end{document}